\documentclass[aps,prl,twocolumn,reprint,groupedaddress]{revtex4-1}

\usepackage{graphicx}
\usepackage{dcolumn}
\usepackage{bm}
\begin{document}
\preprint{Preprint stuff}

\title{Observation of a strong coupling effect on electron-ion collisions in ultracold plasmas}

\author{Wei-Ting Chen, Craig Witte and Jacob L. Roberts}
 
\affiliation{%
Department of Physics, Colorado State University, Fort Collins, CO 80523
}%

\date{\today}

\begin{abstract}

Ultracold plasmas (UCPs) provide a well-controlled system for studying multiple aspects in plasma physics that include collisions and strong coupling effects. By applying a short electric field pulse to a UCP, a plasma electron center-of-mass (CM) oscillation can be initiated. For accessible parameter ranges, the damping rate of this oscillation is determined by the electron-ion collision rate. We performed measurements of the oscillation damping rate with such parameters and compared the measured rates to both a molecular dynamic (MD) simulation that includes strong coupling effects and a Monte-Carlo binary collision simulation designed to predict the damping rate including only weak coupling considerations. We found agreement between the experimentally measured damping rate and the MD result. This agreement did require including the influence of a previously unreported UCP heating mechanism whereby the presence of a DC electric field during ionization increased the electron temperature, but estimations and simulations indicate that such a heating mechanism should be present for our parameters. The measured damping rate at our coldest electron temperature conditions was much faster than the weak coupling prediction obtained from the Monte-Carlo binary collision simulation, which indicates the presence of a significant strong coupling influence. The density averaged electron strong coupling parameter $\Gamma$ measured at our coldest electron temperature conditions was 0.35. 

\end{abstract}

\maketitle

Electron-ion collisions are a fundamental feature of plasmas that determine several plasma properties, such as electron-ion thermalization rates \cite{Dimonte2008}, transport coefficients (diffusion, electric conductivity) \cite{Hinton}, and stopping power considerations that, for instance, influence achievable DT fusion \cite{Zylstra2015,Frenje}. For a weakly coupled plasma, the electron-ion collision rate is given by \cite{Schneider2000}

\begin{eqnarray}
\nu_{ei} = \frac{1}{3}\sqrt{\frac{2}{\pi}} \frac{Z^{2}e^{4}n_{i}}{4 \pi \epsilon_{0}^{2} m_{e}^2 v_{th}^{3}} \ln{\Lambda},
\end{eqnarray}

\noindent where $Z$ is the ion charge number, $e$ is the elementary electron charge, $n_{i}$ is the ion density, $\epsilon_{0}$ is the electric permittivity in vacuum, $m_{e}$ is the mass of an electron, $v_{th}=\sqrt{k_{b}T_{e}/m_{e}}$, and $\ln{\Lambda} = \ln{(C\lambda_{D}/b_{0})}$ is called the Coulomb logarithm, where $\lambda_{D}$ is the Debye screening length, $b_{0}=e^{2}/4\pi \epsilon_{0} k_{b} T$ is the characteristic large angle scattering impact parameter, where $\epsilon_{0}$ is electric permittivity, and $k_{b}$ is Boltzmann constant, and $C$ is a constant, suggested to be 0.765 in Ref. \cite{BPS2005,Dimonte2008,Grabowski2013}.

The presence of the screening length in the collision rate shows collective effects are relevant in a plasma even for individual collisions. This comes about because of a logarithmic divergence in the computed collision rate arising from large impact parameter collisions. The screening in a plasma reduces the influence of such collisions by  screening out the inter-particle Coulomb forces. When the screening length $\lambda_{D}$ is much larger than other  scale lengths such as $b_{0}$ or the typical interparticle spacing given by the Wigner-Seitz radius $a$, the assumptions that go into the derivation of Eq. 1 are valid. For sufficiently cold and dense plasmas, however, $\lambda_{D}$ becomes on the order of $a$ and $b_{0}$, and spatial correlations develop resulting in a more complicated situation. 

The importance of such spatial correlations in a plasma can be characterized by the strong coupling parameter $\Gamma=[b_{0}/(\sqrt{3} \lambda_D)]^{2/3}$ \cite{Ichimaru}, which equals to the ratio between the nearest-neighbor Coulomb energy and the average thermal energy. In our work, strong coupling is considered for the electron component of the plasma, not the ion component as in work elsewhere \cite{Bannasch}. For sufficiently strongly coupled plasmas, Eq. 1 must break down. This can be seen by rewriting Eq. 1 in terms of the electron plasma frequency ($\omega_{p}$) and $\Gamma$ as $\nu_{ei}=\omega_{p} \sqrt{\frac{2}{3 \pi}} \Gamma^{3/2} \ln{(\frac{C}{\sqrt{3}} \Gamma^{-3/2})}$ and noting that for high enough $\Gamma$, the collision rate will be predicted to be unphysically negative.

Strongly coupled plasmas are found in natural \cite{VanHorn1991} and laboratory \cite{Thomas, Tan1995, Ravasio,Murillo2004,Hu,Adams} plasma systems. It is challenging to explore strong coupling effects on electron-ion collisions experimentally due to the difficulty in maintaining a plasma with a sufficiently strongly coupled electron component. However, ultracold plasmas (UCPs) are cold and dense enough to enable such measurements to be conducted. In this work, we describe using an electron oscillation to measure the electron-ion collision rate. The measured rates are in good agreement with molecular dynamics (MD) simulations. In contrast, we found a well-resolved disagreement between the measured rate and predictions based on only weak coupling considerations, making these results the first demonstration of strong coupling influence on electron-ion collision rates in a system free of significant interaction with neutrals. While previous strong coupling theoretical extensions naively indicate 50 $\%$ corrections for our conditions \cite{Dimonte2008,Grabowski2013,Baalrud,Stanton}, we see a factor of 4 increase instead -- a much larger effect that is described in detail below.  

To create our UCP, we first made an $^{85}$Rb magneto-optical trap and then loaded the atoms into an anti-Helmholtz coil magnetic trap. After the atoms were loaded into the magnetic trap, they were transferred to another chamber for plasma creation via two-step photoionization \cite{Chen2016}. Through controlling the wavelength of the photoionizing laser, the initial kinetic energy imparted to the UCP electrons could be controlled. Through adjusting the intensity  of the laser associated with the first step of the two-step photoionization, we can control the number of the electrons and ions via photoionization, and we typically ionize about 5$\%$ of the initial cold atom gas. After photoionization, electrons will immediately leave the UCP until a sufficiently large space charge develops such that the remaining electrons are trapped, forming a plasma. Typical electron and ion temperatures can be as low as a few Kelvins \cite{Killian1999}. The plasma then will expand and fall apart on the order of one hundred $\mu s$, but all the measurements reported here occurred before such expansion was significant.

Our primary experimental signal consisted of measuring the electrons' escape from the UCP, both in response to sequences of applied electric field pulses and as a result of unperturbed UCP evolution. There was an axial 2V/m DC electric field and a 9 G magnetic field applied that helped guide the escaping electrons to the detector, a micro-channel-plate. Typical plasma ion numbers $N_{i}$ were $6.9 \times 10^{4}$ ions with a spatial distribution $n_{i}e^{-r^2/2\sigma^{2}}$, where $r$ is the distance to the center of the plasma, $n_{i}$ is the peak ion density, and $\sigma$ is the characteristic spatial extent, which is about 650 $\mu m$ for our experiments.  

To perform our experiments, we created the UCPs at a higher applied electric field then ramped that field down to 2 V/m so as to operate at a desired charge imbalance (i.e. desired electron to ion number ratio). The chosen high charge imbalance was selected to operate in the regime where electron oscillation damping was predicted to be dominated by electron-ion collisions \cite{Chen2016}. 3.6 $\mu s$ after the plasma was created, we applied a short electric pulse along the DC electric field direction to 'kick' the electrons to initiate the electron center-of-mass (CM) oscillation. During such an oscillation, electric fields are generated that drive electrons with particular velocities and positions out of the UCP. The total number of electrons driven out is linearly proportional to the amplitude of the oscillation\cite{Chen2016}. 

\begin{figure}
\includegraphics{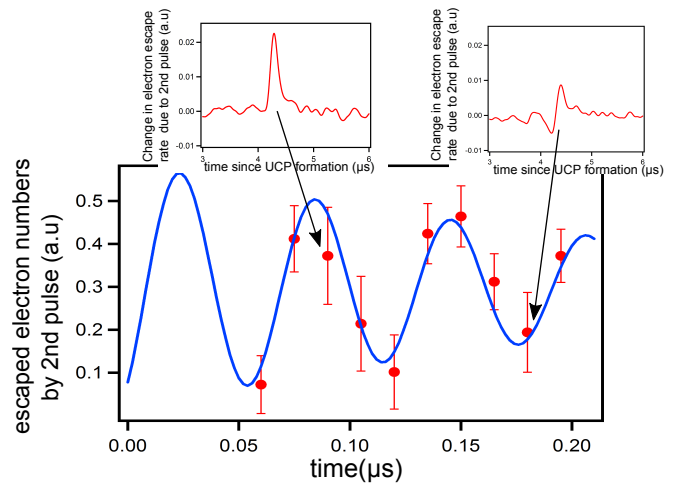}
\caption{A typical experimental data set. The data points show the electron escape as a function of time delay between the initial and second applied electric field pulse. The solid line is the damped cosine wave fit to the data. This set was taken at initial electron kinetic energy $2.26 K \cdot{k_{b}}$. The ion and electron numbers were $5.9 \times 10^{4}$ and $2.7 \times 10^{4}$ respectively. The two insets show the recorded net electron escape signal as detected for a relatively large escape point (left) and a relatively low escape point (right).}
\label{signalillustration}
\end{figure}

This allowed us to measure the oscillation amplitude by applying a second electric pulse after a chosen delay time to modify the amplitude of the oscillation. If the second pulse hits in phase with the induced electron oscillation, then the oscillation amplitude would be increased, which would then drive more electrons out of the UCP. Likewise, when the second pulse was applied with a delay time so that it hit out of phase with the oscillation, the electrons escape due to the oscillation was reduced \cite{Chen2016}. Thus, by measuring the number of the electrons that escaped as a result of the second pulse as a function of the delay time between the two pulses, we mapped out the original electron oscillation amplitude as a function of time. A typical data set from such a measurement set is shown in Fig. \ref{signalillustration}. We performed the measurement at two initial ionization energies : $3.12\times 10^{-23}$J (or $k_{b} \cdot 2.26$ K in temperature equivalent units), and $1.38\times 10^{-24}$J ($k_{b} \cdot 0.1$ K). These values are chosen to be just above ionization threshold on the low side, and not too hot on the high side such that the damping rate would be difficult to resolve. 

To extract the decay rate of the oscillation, we fit the data such as shown in Fig. \ref{signalillustration} to a damped cosine wave. There were complications that we had to deal with in doing so. There existed shot-to-shot variation in particle number, spatial size, and charge imbalance. This effectively added a phase variation that would introduce an additional apparent damping into the data since the oscillation frequency is sensitive to those parameters \cite{Chen2016}. However, the charge imbalance and the total number were measured with high precision for each individual data point, and the size is a function of those two parameters. To mitigate this problem, we introduced correction terms to compensate the variations on a point-to-point basis in the damped cosine fit. We analyzed random simulated data using our analysis protocol to ensure proper fitting of the damping rate and the determination of the associated uncertainties. The resulting damping rates were 3.72$\mu s^{-1} \pm$ 0.79 $\mu s^{-1}$for initial ionization energy $k_{b} \cdot 2.26$ K, and 8.53$\mu s^{-1} \pm$ 1.54 $\mu s^{-1}$ for initial ionization energy $k_{b} \cdot 0.1$ K for our average particle number and spatial size conditions.

In order to determine the predicted electron CM oscillation damping rate given an electron-ion collision rate in, say, Eq. 1, the collision and damping rates need to be related to one another. If the electrons remained in thermal equilibrium during the oscillation (i.e. could be treated hydrodynamically), an analytical relationship can be easily derived. However, the electron-electron collision timescale is on the order of the electron-ion collision timescale, and so the electrons cannot be assumed to be in thermal equilibrium during the damping measurement. In addition, the scattering rates of slower electrons are higher than faster electrons. This leads to non-trivial velocity-space correlations that must be explicitly accounted for. 

In order to take such an effect into account, we developed a numerical model capable of linking any predicted electron-ion collision rate to a predicted oscillation damping rate. In this model, the electrons are tracked as individual particles that interact with one another via Coulomb forces. They are placed in a smooth positive charge background based on the average ion density to model the electron confinement. The electron-ion collisions are modeled with a random collision operator that consists of three parts. First, for an electron moving with velocity $v$, a maximum possible impact parameter $b_{max}$ is computed. Second, the probability of a collision in each timestep $dt$, is calculated as $nv\pi b_{max}^{2} dt$, and random numbers are generated for each electron to see if a collision occurs. Finally, if a collision does happen, the impact parameter is randomly determined and the resulting electron velocity deflection from the collision is applied accordingly and instantaneously. 

We apply all of the usual assumptions typically used in weak-coupled electron-ion collision calculations: Rutherford scattering with a cutoff parameter $b_{max}$ based on $\lambda_{D}$, the binary collision approximation, a substitution of a thermal velocity in the Coulomb logarithm for an individual electron velocity, and an assumption that ions are spatially uncorrelated \cite{Spitzer}. Within these approximations, any electron-ion collision expression translates to a unique random collision operator. Thus, given a model for electron-ion collisions, an electron oscillation damping rate can be calculated. For weak-coupling predictions, we set the collision cross-section to produce collision rates consistent with those implied in the so called BPS stopping power model \cite{BPS2005}. 

\begin{figure}
\includegraphics{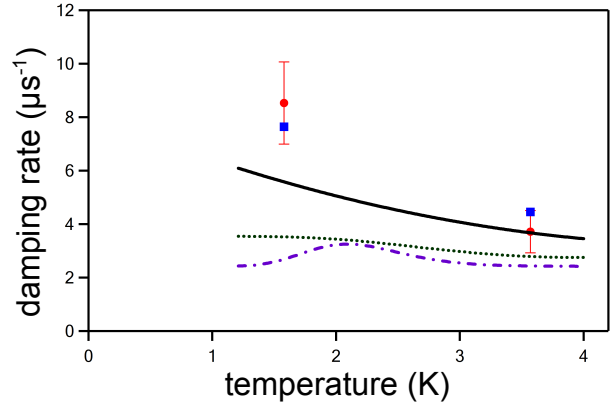}
\caption{Comparisons between measured damping rates and calculations. The red circles are the measured damping rate. The blue squares are MD simulation results. All lines are results from MC binary collision model simulation results (see main text). The dash-dotted purple line is weak coupling result. The green dotted line is using strong coupling extensions from Ref. \cite{Dimonte2008,Grabowski2013,Baalrud}. The black solid curve is the result using a fixed velocity-independent cutoff in addition to account for strong coupling corrections.}
\label{all}
\end{figure}

To see if strong coupling corrections are necessary, we performed calculations around the maximum $\nu_{ei}$ in Eq. 1. We found that the weak-coupling limit damping rate peaked around 2 K with rate 3.3$\mu s^{-1}$ as shown in Fig. \ref{all}. The predicted maximum damping rate is beyond 3 standard deviations below the measured damping rate of our colder temperature data. Thus the weak coupling predictions cannot match our experimental observation for any electron temperature.

To capture the strong coupling effect directly to compare with our data, we performed full MD simulations for both electrons and ions. In other words, we modeled the electron-ion collisions through direct Coulomb force calculations while tracking both electron and ion positions and velocities rather than modeling electron-ion collisions with a random collision operator. A softening parameter for the Coulomb potential was used to address timescale and other problems associated with unlike charges \cite{Oneil}. The MD simulation predicted larger damping rates for parameters where the strong coupling is significant, as expected. By tuning the electron temperature, the predicted damping rate could be matched to our experimentally measured damping rates. The main question is thus whether the implied temperature obtained through this technique is consistent with expectations.

There were several factors that influenced the electron temperature of our UCPs. The first one was the initial ionization energy, which was determined by the photoionization laser wavelength. Other known effects include continuum lowering \cite{Stewart1966}, three-body recombination heating \cite{Robicheaux2002}, disorder induced heating \cite{Murillo2001}, evaporative cooling \cite{Wilson2013_2}, and adiabatic cooling \cite{Robicheaux2002}. Our conditions were chosen to minimize all these factors, and we used the MD simulation to calculate their net contribution. When we did so, we found that the most significant heating arose from a previously unreported mechanism where a DC electric field applied during the formation raises the electron temperature. This heating occurs because the DC electric field accelerates the electrons while the trapping potential due to the ions develops. 

Using MD simulations, we can predict the amount of heating as a function of the initial ionization energy and applied DC electric field. The DC electric field was measured to be 2.0(1) V/m using our electric field calibration procedure described in Ref. \cite{Wilson2013_2}. By running the MD simulation using this measured electric field while initializing the simulation with our experimental conditions, we can compare the measured damping rates to predicted damping rates as shown in Fig. \ref{all}. We found reasonable agreement between the two. The consistency between MD simulation results and experimental measurements gives us confidence that the relevant physics considerations are accounted for in the MD simulation. We therefore used the MD simulation to extract the temperature from the measured damping rate. The temperature of the hotter set of data was determined to be 3.57 K $\pm$ 0.71 K and the colder set to be 1.58 K $\pm$ 0.28 K, corresponding to a density-averaged $\Gamma$ of 0.15 $\pm 0.04$ and 0.35 $\pm 0.08$ respectively. The observation of $\Gamma$=0.35(8) is interesting in light of predictions that $\Gamma$ would not exceed 0.2 \cite{Robicheaux2002} in UCPs. It is not that our results show Ref. \cite{Robicheaux2002} to be incorrect. Rather, at sufficiently low density, the three-body recombination rate becomes slow enough compared to UCP formation time, so that $\Gamma$ can exceed 0.2 \cite{Guthrie2016}.


Although the MD results are in agreement with the measured data, it is interesting to compare predicted strong coupling extensions using our MC binary collision code to our experimental results. These extensions have been theoretically developed in multiple contexts including electron-ion temperature equilibration, stopping power, effective potential, and transport and diffusion calculations  \cite{Dimonte2008,Grabowski2013,Baalrud,Stanton}. However, given that the oscillating CM electron velocity in these experiments is less than the typical electron thermal velocity (at least for most of the oscillation period), the range of comparison for these different approaches in the relevant parameter space is expected to be dominated by binary electron-ion collisions. Not surprisingly, these theories produce expressions that involve  a modification of the Coulomb logarithm. For the range of $\Gamma$ in our work, a modification consistent with the predictions in Ref. \cite{Dimonte2008,Grabowski2013,Baalrud} is to set $\ln{\Lambda}=\ln{(1+0.765\lambda_{D}/b_{0})}$ in Eq. 1. We concentrated on comparisons to these theories because they are consistent with one another and are referenced directly to classical MD simulations that involve assumptions that look to be well justified for UCPs. Through applying these extensions to our MC binary collision calculation, we can compute the resulting modification of weak-coupling CM damping rate. For our colder conditions, the predicted damping rate is 3.41 $\mu s^{-1}$. This is an improvement on the weak-coupling-only prediction, but still not consistent with our measurement or associated MD simulation. Thus, a straightforward application of the implied collision rates in Ref. \cite{Dimonte2008,Grabowski2013,Baalrud} fails to match our observations. 

A natural implication is that one or more of the standard assumptions that were included in MC binary collision calculation are violated. The most suspect is the thermal velocity substitution approximation, and we investigated what happens if that approximation is modified. We did so by removing the approximation completely while still requiring consistency with the predictions in Ref. \cite{Dimonte2008,Grabowski2013,Baalrud}. By just removing this assumption, more than half the gap between the strong coupling predictions with the assumption (green dotted line in Fig. \ref{all}) and MD results was closed. While we were altering the nature of the cutoff ($b_{max}$) with velocity, we explored related impacts further by including dynamic screening that scales with each electron's velocity, but if consistency with Ref. \cite{Dimonte2008} is maintained, the change in predicted damping rate from such a dynamic screening is less than a few percent. Despite achieving significant improvement by introducing the modifications above, there is still a difference that remained with respect to MD simulation/experimental results.


Thus, it seems that one or more other remaining assumptions are also violated. We examined the possible influence of ion-ion spatial correlations by increasing their mass substantially in the MD code to greatly slow down any correlation formation. Electron-ion correlation influence was also examined by MD simulation using like charge ions and electrons within a smoothed neutralizing background to see if unlike charge effects were significant. The changes in damping rate in both cases are found to be less than few percent- not enough for agreement. Therefore, it appears that the Rutherford scattering or binary collision approximation, or both, are not valid. Investigation of the breakdown in these assumptions is more complicated than relaxing other assumptions and is to be the subject of planned future work. In any case, we show that the standard electron-ion collision approximations are problematic and need to be treated with care when strong coupling is relevant.

In conclusion, we experimentally measured a strong coupling influence on the electron-ion collision rate in a UCP. Our experimental results were consistent with molecular dynamics modeling of our system. We report a measured electron strong coupling parameter as large as $\Gamma = 0.35(8)$, which demonstrates that experimental conditions at low density can achieve greater value of $\Gamma$ than predicted in \cite{Robicheaux2002}, consistent with other predictions \cite{Pohl,Robicheaux2011}. In addition, we identify a previously unreported heating mechanism that occurs due to the presence of a DC electric field during UCP formation. Under the listed conditions, the necessity of strong coupling corrections is not surprising, but the size of correction is larger than expected from other theories \cite{Dimonte2008,Grabowski2013,Baalrud,Stanton}, and if typical assumptions are applied, it is not possible to obtain simultaneous agreement between our data and these theories. We found improvements if the standard practice of replacing velocity terms with temperature terms in the Coulomb logarithm is relaxed, but that change alone was not enough to produce agreement. This likely indicates breakdown other assumptions, and further investigation of the validity of assumptions of Rutherford scattering and binary collision are called for, and are subjects for future investigations.

\begin{acknowledgments}
We would like to thank for the Air Force Office of Science Research, Grant No. FA 9550-12-1-0222 for supporting this research.
\end{acknowledgments}

\end{document}